# Layer dependence of geometric, electronic and piezoelectric properties of SnSe

Wuzhang Fang,[1] Li-Chuan Zhang,[2] Guangzhao Qin,[2,3] Qing-Bo Yan,[2,*] Qing-Rong Zheng,[1,*] Gang Su[1,*]

[1]School of Physical Sciences, University of Chinese Academy of Sciences, Beijing 100049, China

[2]College of Materials Science and Opto-Electronic Technology, University of Chinese Academy of Sciences, Beijing 100049, China

[3]Institute of Mineral Engineering, Division of Materials Science and Engineering, Faculty of Georesources and Materials Engineering, RWTH Aachen University, Aachen 52064, Germany

**Corresponding Author**

*Email: (Q.-B.Y.) yan@ucas.ac.cn.

*Email: (Q.-R.Z.) qrzheng@ucas.ac.cn.

*Email: (G.S.) gsu@ucas.ac.cn.

**Abstract**

By means of first-principles calculations, we explore systematically the geometric, electronic and piezoelectric properties of multilayer SnSe. We find that these properties are layer-dependent, indicating that the interlayer interaction plays an important role. With increasing the number of SnSe layers from 1 to 6, we observe that the lattice constant decreases from 4.27 Å to 4.22 Å along zigzag direction, and increases from 4.41 Å to 4.51 Å along armchair direction close to the bulk limit (4.21 Å and 4.52 Å, respectively); the band gap decreases from 1.45 eV to 1.08 eV, approaching the bulk gap 0.95 eV. Although the monolayer SnSe exhibits almost symmetric geometric and electronic structures along zigzag and armchair directions, bulk SnSe is obviously anisotropic, showing that the stacking of layers enhances the anisotropic character of SnSe. As bulk and even-layer SnSe have inversion centers, they cannot exhibit piezoelectric responses. However, we show that the odd-layer SnSe have piezoelectric coefficients much higher than those of the known piezoelectric materials, suggesting that the odd-layer SnSe is a good piezoelectric material.

**Introduction**

Two-dimensional (2D) materials including graphene, hexagonal boron-nitride (h-BN), $MoS_2$ and black phosphorus (BP), etc. have led to a lot of studies on their novel properties and potential applications in functional devices [1-4]. Compared to the single atomic layer of 2D materials, the multi-layers of 2D materials are easier to fabricate and more common in experiments, which have shown fruitful layer-dependent features in electronic, vibrational and optical responses due to the additional interlayer interaction [5-8]. The multilayer 2D materials also have great potential applications. For example, multilayer $MoS_2$ and BP have been used to make high performance field-effect transistors, which have rather high mobility of 700 $cm^2 V^{-1} s^{-1}$ and 1000 $cm^2 V^{-1} s^{-1}$ [9, 10], respectively, demonstrating their potential in future nanoelectronic devices.

SnSe is a layered material, which has a structure similar to BP and has wide potential applications in photovoltaic and thermoelectric devices [11-13]. Recently, the synthesis of 2D SnSe has been reported [14, 15]. Monolayer SnSe is disclosed to be a semiconductor with an indirect gap (1.45 eV), a small Young's modulus, a low lattice thermal conductivity and a high hole mobility [16, 17]. It is also found that the piezoelectricity of monolayer SnSe is giant [18] and the piezoelectric coefficient $d_{11}$ is about two orders of magnitude larger than those of other frequently used and 2D piezoelectric materials, such as α-quartz, $MoS_2$, h-BN and graphene with absorption atoms [19-26]. However, the physical properties of multilayer SnSe are sparsely investigated. How does the number of layers affect the properties of SnSe? In particular, the piezoelectricity of bulk SnSe is absent while the monolayer SnSe is giant. What is the case for multilayer SnSe? These issues deserve to pay more attention.

In this paper, we systematically investigate the geometric, electronic and piezoelectric properties of multilayer SnSe from 1-6 layer by means of first-principles density functional theory (DFT) calculations. We find that with increasing the number of layers of SnSe, the lattice constant increases along zigzag direction while decreases along armchair direction. The electronic band structure is very symmetric for monolayer SnSe, but it becomes more and more asymmetric as the layer number increases. The piezoelectric coefficients are found nonzero only in odd-layer SnSe, which are much larger than those of known piezoelectric materials, and decrease with increasing the layer number. We also propose several ways to enhance the piezoelectric effects in multilayer SnSe to make it more practical.

**Calculation methods**

All first-principles DFT calculations are performed using the plane wave basis set, the Perdew-Burke-Ernzerhof (PBE) [27] of generalized gradient approximation (GGA) for exchange-correlation potential and the projector augmented wave (PAW) method [28, 29] as implemented in the Vienna *ab initio* simulation package (VASP) [30]. The kinetic energy cutoff for plane wave basis is set to 600eV. A Monkhorst-Pack mesh [31] of $13 \times 13 \times 1$ is used to sample the Brillouin zone and the energy convergence threshold is set to $10^{-6}$ eV. A large vacuum layer no smaller than 15 Å is used. The geometries are fully optimized until the residual force on each atom is smaller than $10^{-3}$ eV Å$^{-1}$ and van der Waals interactions are considered using the optB88 exchange functional (optB88-vdW) [32, 33]. The modified Becke-Johnson (mBJ) method [34] is adopted to calculate electronic band structures. In the calculation of phonon dispersion, a $5 \times 5 \times 1$ supercell and a Monkhorst-Pack mesh of $1 \times 1 \times 1$ are used to calculate the second order interatomic force constants (IFCs) within the finite displacement method. Then the phonon dispersion is obtained using the PHONOPY package [35].

**Geometric structures**

SnSe has a hinge-like layered structure along armchair direction, similar to BP [36]. The structure of bilayer SnSe is shown in Fig. 1. The space group is $P2_1/m$ for bilayer SnSe and $Pmn2_1$ for monolayer SnSe. In bilayer SnSe, one layer is translated relative to its neighboring layer, resulting in a centrosymmetric symmetry. This centrosymmetric symmetry makes bilayer SnSe non-piezoelectric. The structures of multilayer SnSe from monolayer to six layers are fully optimized and the lattice constants are summarized in Table 1 and Fig. 2a. The space groups of odd- and even-layer SnSe are listed in Table1. The main difference between odd- and even-layer SnSe is that the even-layer SnSe has an inversion center while the odd-layer SnSe does not. As shown in Fig. 2a, the lattice constant decreases along zigzag direction while increases along armchair direction with increasing the layer number, resulting in an increasing anisotropy in geometry, which may have consequences on physical properties. The change of lattice constant from monolayer to bulk is 0.06 Å along zigzag direction and 0.11 Å along armchair direction, but there is a big change between the monolayer and bilayer, which is attributed to the interlayer interaction in bilayer (0.59 eV/cell). This is in contrast to the multilayer BP, where the increase of layer number gives rise to a decreasing anisotropy in geometry [37]. The reason may be from the difference of interlayer interactions in both materials [38].

**Phononic and electronic properties**

The phonon dispersions of monolayer and bilayer SnSe are shown in Fig. 3. No imaginary frequency is found. In one unit cell, there are four atoms in the monolayer and eight atoms in the bilayer. Therefore, there are 12 phonon branches in the monolayer and 24 branches in the bilayer. The phonon dispersion of bilayer resembles the one of monolayer, and most of the branches in the bilayer can be viewed as the ones split from the monolayer. The frequencies of dispersion in the bilayer are depressed compared with the ones in the monolayer. The maximal frequency of the monolayer and bilayer is 5.2 THz and 5.0 THz, respectively. The phonon group velocity can be obtained from the slope of the longitudinal acoustic branch near the Gamma point, which has the values along $G-X$ and $G-Y$ 3.0 km s$^{-1}$ and 2.7 km s$^{-1}$ for monolayer, and 2.9 km s$^{-1}$ and 2.3 km s$^{-1}$ for bilayer, respectively.

The interlayer interaction also affects the electronic structures of multilayer SnSe. The calculated band structures from monolayer to six layers are summarized in Table 1 and Fig. 2b. One may see that the multilayer SnSe is an indirect semiconductor. The band gap decreases from 1.45 eV of monolayer to 1.08 eV of six layers (with mBJ functional), showing that the band gap is decreasing with increasing the layer number. The gap of six layers of SnSe is close to the bulk limit. This decreasing trend with the increasing layer number in the band gap of multilayer SnSe is similar to the case of multilayer BP [37]. The electronic band structure is very symmetric for the monolayer SnSe but becomes more and more asymmetric as the number of layers increases, which is consistent with the increasing anisotropy in geometry.

**Piezoelectric properties**

To describe the piezoelectricity, two third-rank piezoelectric tensors $d_{ijk}$ and $e_{ijk}$ are used, which are defined by:

$$d_{ijk} = \left(\frac{\partial P_i}{\partial \sigma_{jk}}\right)_{E,T} = \left(\frac{\partial \varepsilon_{jk}}{\partial E_i}\right)_{\sigma,T}$$

$$e_{ijk} = \left(\frac{\partial P_i}{\partial \varepsilon_{jk}}\right)_{E,T} = -\left(\frac{\partial \sigma_{jk}}{\partial E_i}\right)_{\varepsilon,T}$$

$P_i$, $E_i$, $\sigma_{jk}$ and $\varepsilon_{jk}$ denote the polarization, electric field, stress tensor and strain tensor, respectively, where $i,j,k \in (1,2,3)$ correspond to x, y, and z directions. From the definition, the d-tensor says that an applied electric field can lead to a responded strain, and the e-tensor means that an applied strain may bring a change of polarization. The piezoelectric d-tensor and e-tensor are connected by a fourth-rank elastic stiffness tensor $C_{mnjk}$:

$$e_{ijk} = d_{imn} C_{mnjk}$$

With Voigt notation, the piezoelectric tensor can be represented by a matrix. The odd-layer SnSe has a point group of mm2 ($C_{2v}$) while the even-layer SnSe has a point group of 2/m ($C_{2h}$). There is an inversion center in even-layer SnSe, so the piezoelectric coefficients in the even-layer SnSe should be zero by the constraint of symmetry [39]. Due to the restriction of symmetry of point group mm2, in the odd-layer SnSe, there are only several non-zero piezoelectric coefficients of e-tensor, which are $e_{21}$, $e_{22}$, $e_{23}$, $e_{14}$ and $e_{35}$ (the same for d-tensor). Since these are 2D materials, we do not consider the change of strain or polarization along z direction, and all coefficients with the index of 3 can be ignored. $e_{14}$ ($d_{14}$) can also be neglected as it is hard to measure in experiments.

We use the density functional perturbation theory (DFPT) [40] and finite difference methods [41] as implemented in VASP code to calculate the piezoelectric e-tensor and the elastic stiffness tensor from monolayer to six-layer SnSe, respectively. Then we transform the values into 2D form by multiplying the lattice constant of z direction of the unit cell. Particularly, in this situation, the *P* and *C* represent the surface polarization and planar elastic stiffness tensor, respectively. We then use the relations below to get the piezoelectric d-tensor:

$$e_{21} = d_{21}C_{11} + d_{22}C_{21}$$

$$e_{22} = d_{21}C_{12} + d_{22}C_{22}$$

The results are collected in Table 1 and Fig. 4. The piezoelectric coefficients off monolayer SnSe are in the same order of magnitude as in the previous calculation [18]. In Fig. 4a, we can see that the planar elastic stiffness constants increase linearly as the number of layers increases. It is easy to understand that, if we take each layer as a spring, the total spring constant of the integrated parallel springs is proportional to the number of layers. The piezoelectric e-coefficients decrease slowly in the odd-layer SnSe with increasing the layer number, while the ones in the even-layer SnSe are zero (Fig4. c), which is consistent with the constraint of symmetry of space group. This result can be manifested in Fig. 4b. In the even-layer SnSe, due to the inversion symmetry, the accumulated charges of every neighboring two layers bear opposite signs and therefore cancel, resulting in the absence of polarization. In the odd-layer SnSe, the magnitude of piezoelectric coefficient $d_{21}$ decreases from 47.7 in monolayer to 8 in five-layer and $d_{22}$ decreases from 158.2 in monolayer to 27 in five-layer (Fig. 4d). Such a dramatic change mainly arises from the increase of elastic stiffness constants. The slowly decreasing piezoelectric e-coefficients play a secondary role. Compared to the measurements of the piezoelectric coefficients of $MoS_2$ in the experiments [25, 26], the trend that the piezoelectric coefficients decrease as the number of layers increases is similar. Although the piezoelectric d-coefficients decrease dramatically from monolayer to five-layer (about five-fold), the magnitude of d-coefficients in five-layer SnSe is still large compared to the ones in α-quartz and 2D piezoelectric materials (Fig. 4d) [19, 20, 23].

The fact that the piezoelectricity exists only in the odd-layer SnSe and the piezoelectric coefficients decrease with the increase of the layer number may refrain from the application of multilayer SnSe as a piezoelectric material. The key to overcome this difficulty is to break the inversion symmetry in multilayer SnSe. Based on this consideration, there are three possible ways that may enhance the piezoelectric effects of multilayer SnSe: (1) making a substitute doping in multilayer SnSe; 2) twisting or translating one SnSe layer relative to its neighboring layer to break the inversion symmetry; 3) using monolayer SnSe and other 2D monolayer material to make a stacking superlattice.

**Conclusion**

In summary, we have systematically investigated the geometric, electronic and piezoelectric properties of multilayer SnSe by means of first-principles calculations. All these properties are found to be layer-dependent, showing that the interlayer interaction plays an important role. As the number of SnSe layers increases from 1 to 6, we found that the lattice constant decreases from 4.27 Å to 4.22 Å along zigzag direction, and increases from 4.41 Å to 4.51 Å along armchair direction, close to the bulk limit (4.21 Å and 4.52 Å). The maximal frequency of phonon dispersion of the monolayer and bilayer SnSe are 5.2 THz and 5.0 THz, respectively. The phonon group velocities along $G-X$ and $G-Y$ directions are 3.0 kms$^{-1}$ and 2.7 km s$^{-1}$ for monolayer and 2.9 km s$^{-1}$ and 2.3 km s$^{-1}$ for bilayer, respectively. Multilayer SnSe are found to be indirect semiconductors, whose band gap decreases from 1.45 eV in monolayer to 1.08 eV in six layers.

The electronic structure of monolayer SnSe is very symmetric but it becomes more and more asymmetric as the number of layers increases. The piezoelectric coefficients are found nonzero only in odd-layer SnSe, which decrease as the number of layers increases but are still much higher than those in other piezoelectric materials, suggesting that the odd-layer SnSe are promising piezoelectric materials. To make it easily accessible, several ways to make SnSe more practical in piezoelectric application are suggested. This present study gains an insight into designing novel high efficiency piezoelectric materials.

**Acknowledgement**

The authors thank Prof. Zhen-Gang Zhu and Zheng-Chuan Wang of UCAS for helpful discussions. All calculations are performed on Supercomputer Center of Chinese Academy of Sciences and MagicCube (DAWN5000A) in Shanghai Supercomputer Center, China. This work is supported in part by the NSFC (Grant No. 11004239, No. 11474279), the MOST (Grant No. 2012CB932901 and No.2013CB933401) of China, and the fund from CAS.

**Figures**

**Figure 1**

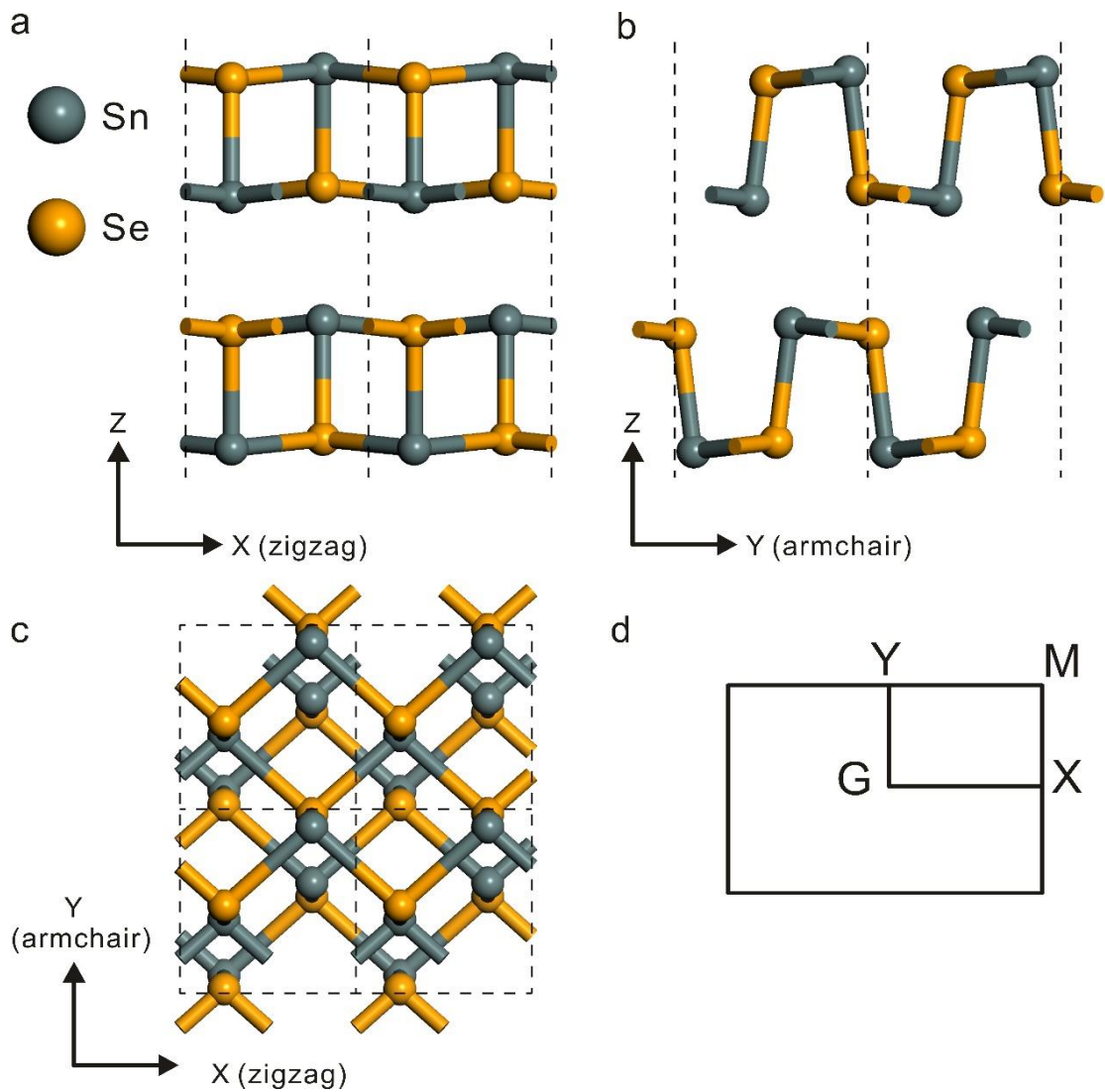

**Fig. 1** Side view along zigzag direction (a) and armchair direction (b) of bilayer SnSe. (c) Top view of bilayer SnSe. (d) The Brillouin zone of multilayer SnSe in calculating the electronic and phononic band structures.

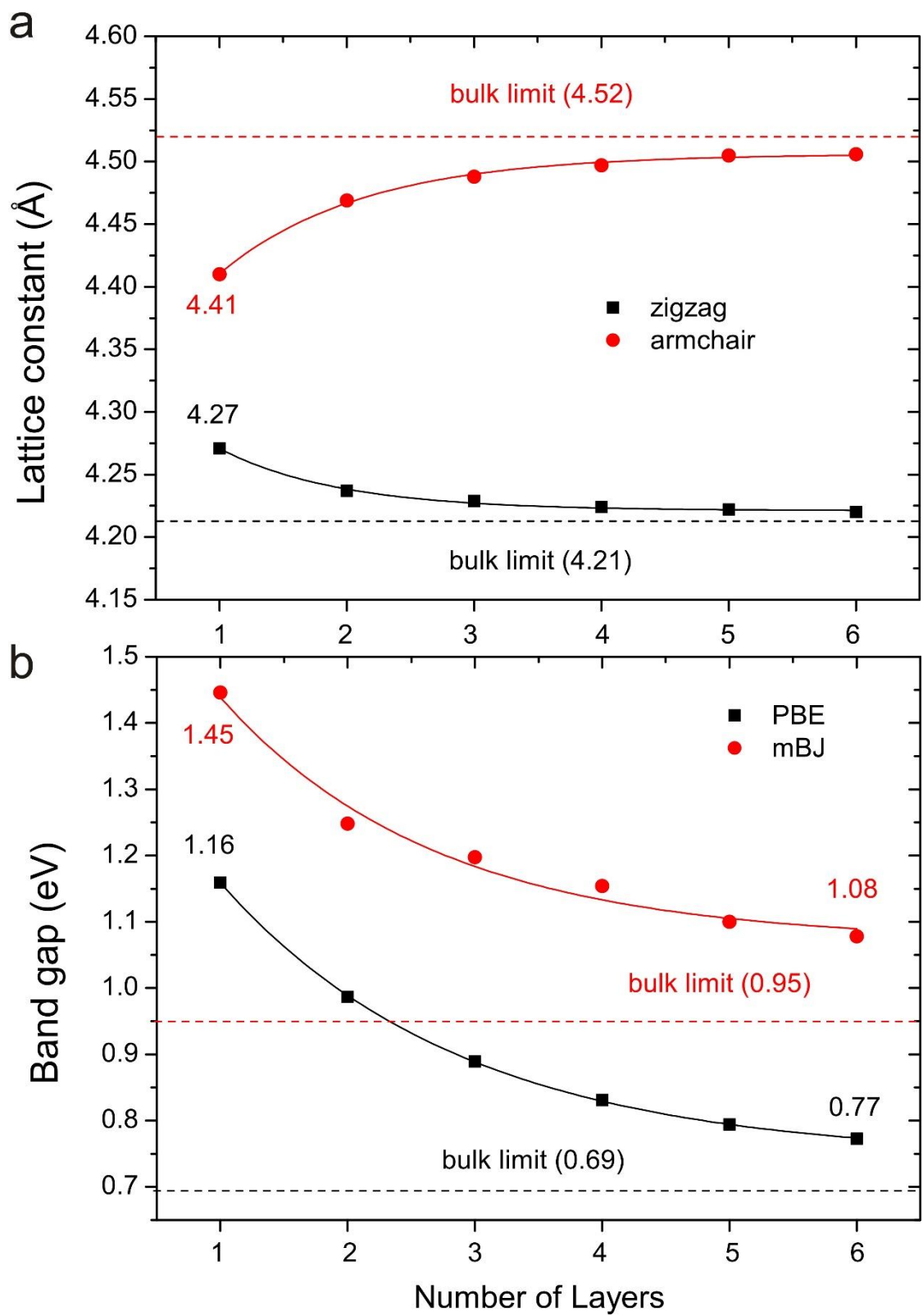

**Fig. 2** (a) The lattice constants of multilayer SnSe (from monolayer to 6 layers) and bulk along zigzag and armchair directions. The dash line represents the bulk limit of lattice constants along zigzag (black) and armchair (red) directions. The solid line represents the fitted curve of lattice constants by an exponential grow (decay). (b) The electronic band gaps of SnSe from monolayer to six layers and bulk calculated with PBE and mBJ functional. The dash line represents the bulk limit of band gaps. The solid line represents the fitted curve of band gaps of SnSe from monolayer to six layers by an exponential decay.

**Figure 3**

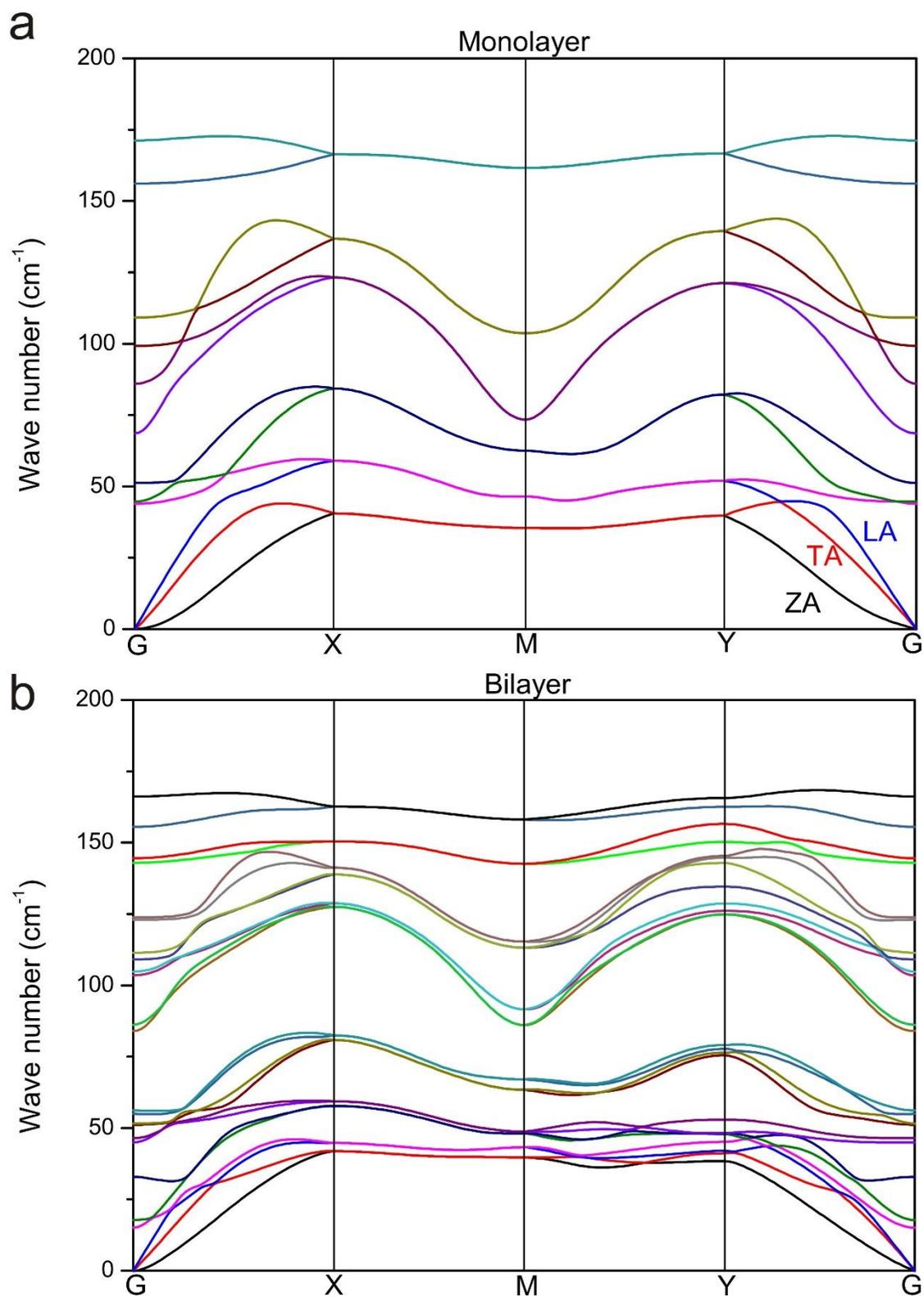

**Fig. 3** The phonon dispersion of (a) monolayer and (b) bilayer SnSe.

**Figure 4**

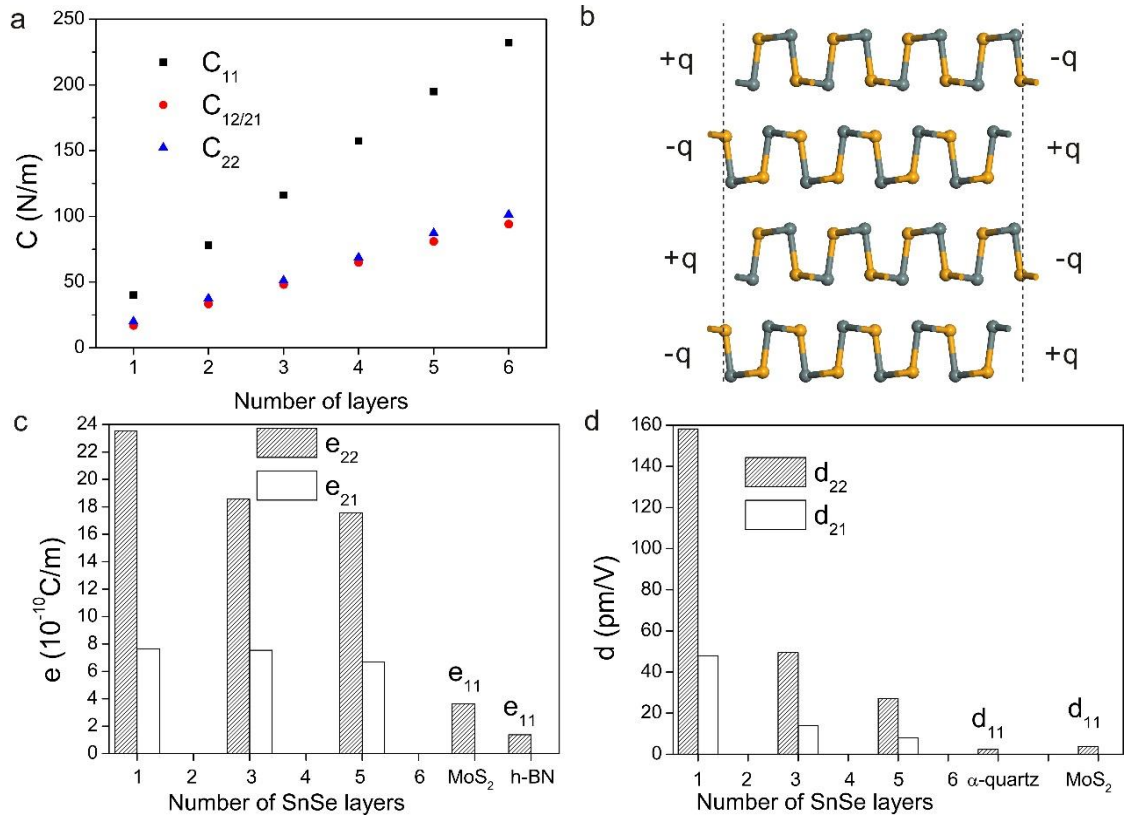

**Fig 4.** (a) The planar elastic stiffness constants, (c) piezoelectric e-coefficients and (d) piezoelectric d-coefficients of multilayer SnSe from monolayer to six layer. (b) Schematic depiction in explaining the absence of piezoelectricity in even-layer SnSe. The piezoelectric coefficients of α-quartz, $MoS_2$ and h-BN are obtained from references [19, 20, 23].

Tables

Table 1

| Number of layers | Space group | Lattice constants Å | | Band gap eV | | $C_{11}$ N/m | $C_{12}/C_{21}$ N/m | $C_{22}$ N/m | $e_{21}$ $10^{-10}$ C/m | $e_{22}$ $10^{-10}$ C/m | $d_{21}$ pm/V | $d_{22}$ pm/V |
|---|---|---|---|---|---|---|---|---|---|---|---|---|
| | | zigzag | armchair | PBE | mBJ | | | | | | | |
| 1 | $Pmn2_1$ | 4.27 | 4.41 | 1.16 | 1.45 | 40.4 | 17 | 20 | 7.62 | 23.54 | -47.7 | 158.2 |
| 2 | $P2_1/m$ | 4.24 | 4.47 | 0.99 | 1.25 | 78.1 | 33.3 | 37.4 | 0 | 0 | 0 | 0 |
| 3 | $Pmn2_1$ | 4.23 | 4.49 | 0.89 | 1.20 | 116.2 | 48.2 | 51.2 | 7.54 | 18.55 | -14 | 49.4 |
| 4 | $P2_1/m$ | 4.22 | 4.50 | 0.83 | 1.15 | 157.4 | 65 | 68.7 | 0 | 0 | 0 | 0 |
| 5 | $Pmn2_1$ | 4.22 | 4.51 | 0.79 | 1.1 | 194.9 | 81 | 87.3 | 6.68 | 17.54 | -8 | 27 |
| 6 | $P2_1/m$ | 4.22 | 4.51 | 0.77 | 1.08 | 232.2 | 94.1 | 101.3 | 0 | 0 | 0 | 0 |

**Table 1** The space groups, lattice constants, electronic band gaps, planar elastic stiffness constants, piezoelectric e-coefficients and d-coefficients of multilayer SnSe from 1 to 6 layers.